\documentclass[sigconf]{acmart}

\AtBeginDocument{%
  \providecommand\BibTeX{{%
    \normalfont B\kern-0.5em{\scshape i\kern-0.25em b}\kern-0.8em\TeX}}}

\setcopyright{acmcopyright}
\copyrightyear{2022}
\acmYear{2022}
\acmDOI{XXXXXXX.XXXXXXX}

\acmConference[FDG`22]{Foundations of Digital Games}{2022}{Athens, Greece}
\acmPrice{15.00}
\acmISBN{978-1-4503-XXXX-X/18/06}

\begin{document}

\title{The Dark Souls of Archaeology: Recording Elden Ring}


\author{Florence Smith Nicholls and Michael Cook}
\affiliation{%
  \institution{Queen Mary University of London}
  \city{London}
  \country{UK}}
 \email{{florence,mike}@knivesandpaintbrushes.org}


\renewcommand{\shortauthors}{Smith Nicholls and Cook}



\begin{abstract}
Archaeology can be broadly defined as the study and interpretation of the past through material remains. Videogame worlds, though immaterial in nature, can also afford opportunities to study the people who existed within them based on what they leave behind. In this paper we present the first formal archaeological survey of a predominantly single-player game, by examining the player-generated content that is asynchronously distributed to players in the videogame \textit{Elden Ring}. We report on our preparation for the study, our findings at the two sites we surveyed, and the lessons learned both specifically about Elden Ring's community and, more broadly, about the nature of archaeological surveying within videogames.
\end{abstract}

\maketitle


\begin{figure}[t] 
\includegraphics[width=0.8\columnwidth]{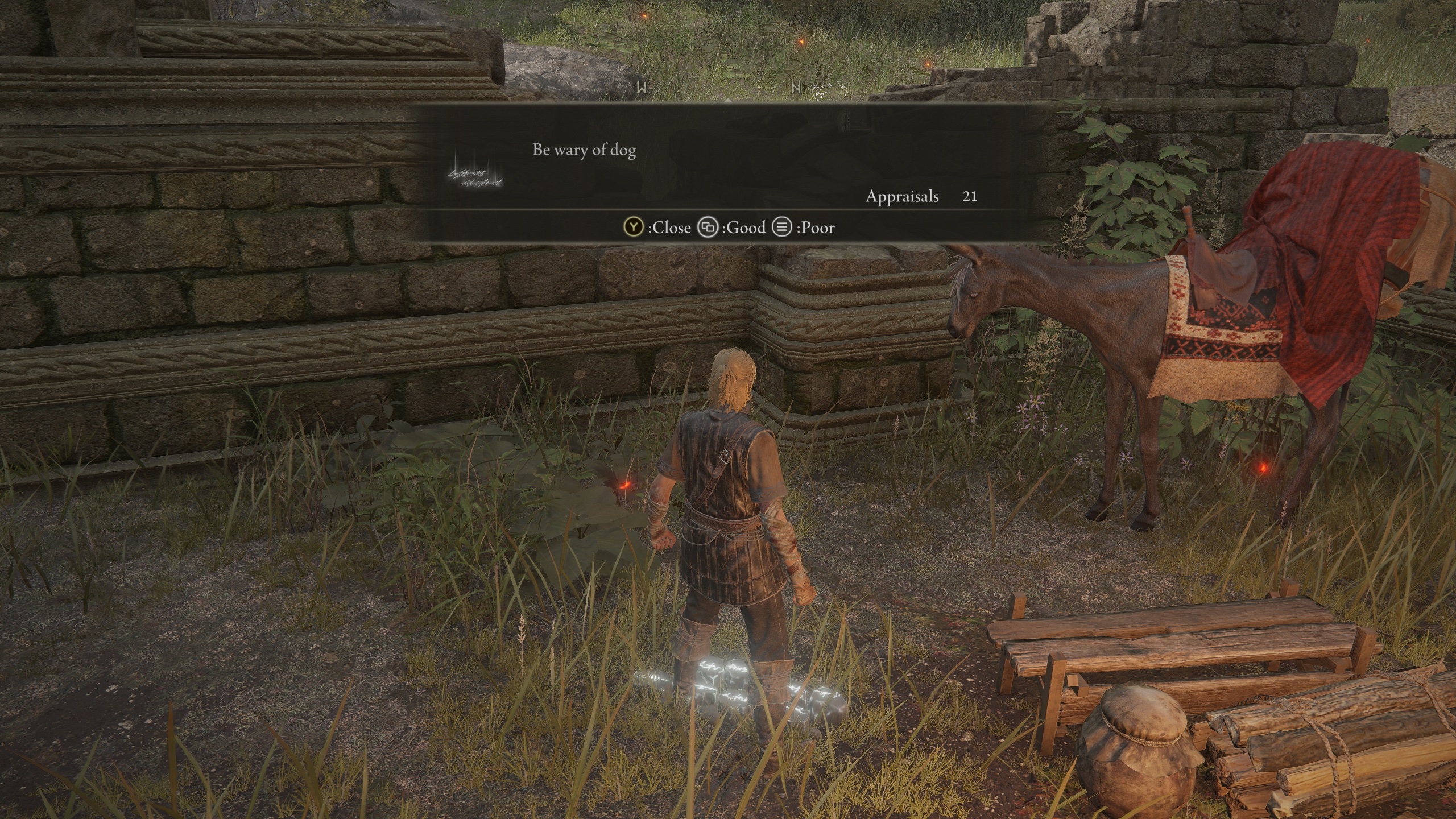}
\caption{A player-written message. It reads `Be wary of dog'.}
\label{fig:message-example}
\end{figure}

\begin{figure}[t] 
\includegraphics[width=0.8\columnwidth]{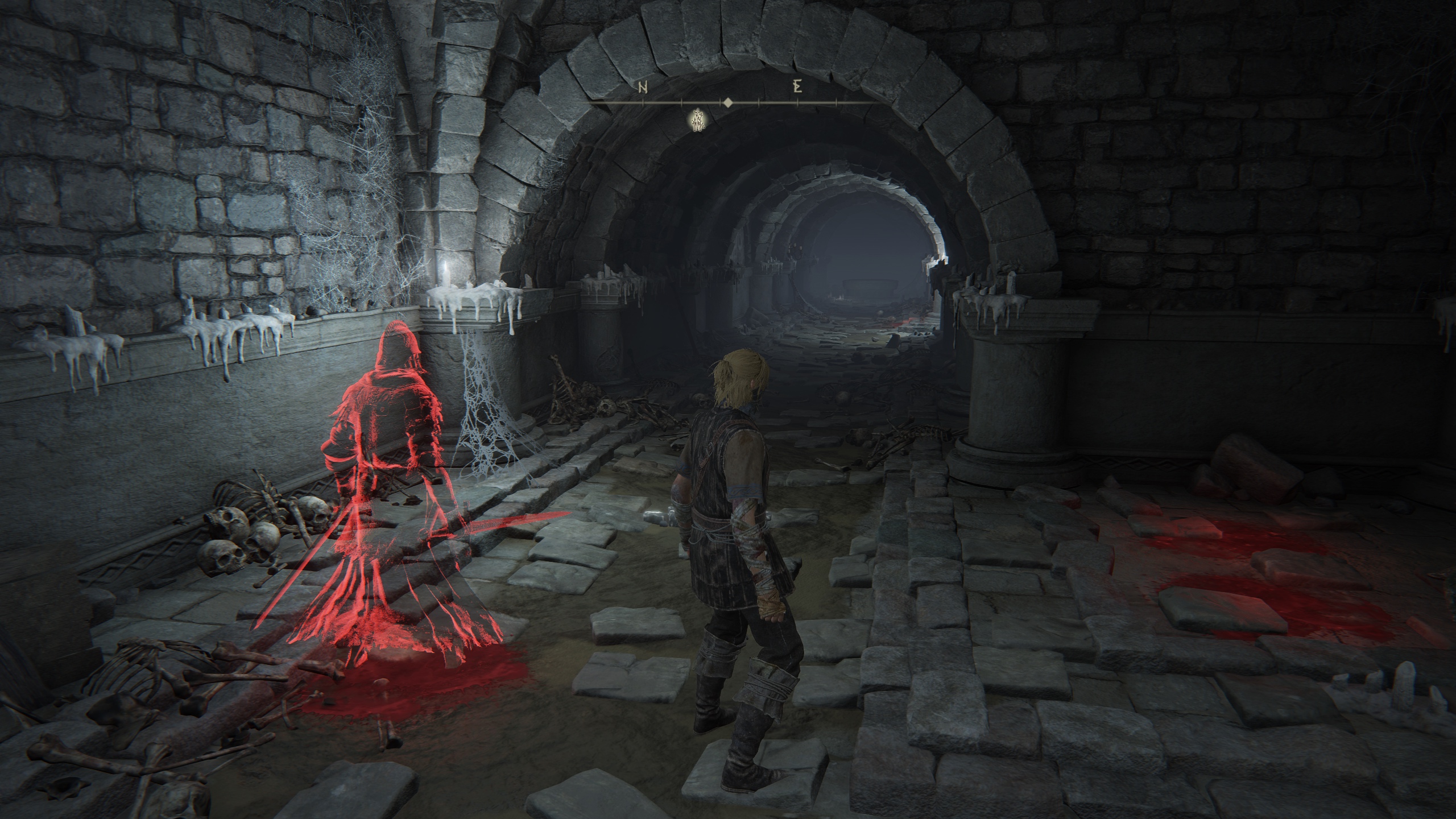}
\caption{Several bloodstains in a dungeon.}
\label{fig:message-example}
\end{figure}

\section{Introduction}
Many different fields of study aim to understand human behaviour. In the study of games, researchers have employed a wide variety of techniques, from simulation and modelling \cite{pfau2020} to surveys and interviews \cite{gow2010}, or simply recording player activity in-game \cite{melhart2021}. Many of these studies rely on an objective ground truth recording of player activity, such as the ability to directly ask a player questions via interview, or extract exact player behaviour data through key logging. In some cases, however, we have evidence of player behaviour through, for example, user-generated content, but no way to access the history of this activity or reach out to the player to ask them questions about what they did, and why.

The field of archaeology also aims to understand human behaviour and activity, but through the study of material remains. Archaeologists are faced with a similar dilemma to the one described above: the people they wish to study are usually not available for questioning, and the remains they have left behind do not tell a complete story, nor were they left behind with the intention of being studied. Instead, archaeologists must interpret these partial material remains and contextualise them within wider research.

For games with user-generated content, then, archaeology offers a new lens through which to study player behaviour. By viewing the remains of player activity as analogous to archaeological remains, we can employ the techniques of an archaeologist to theorise about the behaviour and intentions of both the players whose remains we find, and the wider community they were a part of. This idea of applying archaeological techniques to videogame worlds is one example of an interdisciplinary field known as \textit{archaeogaming}. Applying archaeological methods to games can help reveal insights into player behaviour, taking advantage of decades of archaeological research, and feed back new insights and research questions to the archaeological community.


In this paper we report on an archaeological survey conducted at two sites in the recently-released action-adventure game \textit{Elden Ring}. Elden Ring utilises a form of asynchronous multiplayer, meaning players leave information behind, both intentionally and unintentionally, that can be found by other players later on. We conducted a survey in the starting area of Elden Ring, Limgrave, and we report on the nature of the messages and player death recordings found at both sites, offering a qualitative analysis of the types of activity in these places and what this tells us about the role these features play in the overall game design. We also use this as an opportunity to provide a meta-perspective on the nature of conducting archaeological surveys of transient artifacts within a game environment, as a contrast to prior work in the area.

The remainder of this paper is organised as follows: in section \ref{sec:background} we provide background on Elden Ring and archaeogaming; in section \ref{sec:methodology} we outline our methodology; in section \ref{sec:results} we present an overview of our data and findings; in section \ref{sec:discussion} we discuss our results; finally in section \ref{sec:futurework} we point to future research and conclude in section \ref{sec:conclusions}.

\section{Background} \label{sec:background}

\subsection{Elden Ring}
\textit{Elden Ring} is a 2022 action-adventure game developed by From Software. Although technically a new IP, Elden Ring is a continuation of a series of similar games made by the developer that are connected in mechanics, themes and aesthetics, even though they do not all share the same narrative and setting. This collection of games is sometimes referred to as the `Soulsborne' series, as a portmanteau of the Demon's Souls/Dark Souls series and Bloodborne, however the collection has since been extended with Sekiro: Shadows Die Twice and now Elden Ring, and arguably stretches earlier than Demon's Souls back to include games such as the King's Field series, Shadow Tower and Armored Core: For Answer. 

Soulsborne games, particularly post-Demon's Souls, share a number of common features. These include a variant of permadeath in which the player loses the currency they are holding on death\footnote{This currency can be recovered if they make it back to their corpse without dying, a so-called \textit{corpse run}.}; an emphasis on large, challenging boss fights; and many complex, arcane or hard-to-find secrets. There are also many thematic and narrative links between the games: the corrupting influence of power; the perpetuation of cycles and traditions; and how legacy and history motivate people. Another key connecting feature present in every From Software game from Demon's Souls onwards is a specific form of asynchronous multiplayer interaction.

However, there are some key divergences between Elden Ring and other Soulsborne games. Apart from its open world setting, Elden Ring has a much larger player base. By March 16th 2022, Elden Ring had sold more than 12 million copies after only a few weeks, compared to Dark Souls 3 which has still only sold 10 million copies since 2016 \cite{peppiatt2020}. According to SteamDB, Elden Ring reached a high of over 950,000 concurrent players on March 3rd 2022 \cite{steamdb2022}.

We decided to undertake an archaeological survey of Elden Ring not because it is popular or financially successful, but because it is a particularly active site of player activity in March 2022. Essentially, we wanted to take a snapshot of the game at this peak of popularity which will likely wane over time.

\subsection{Asynchronous Multiplayer}
Soulsborne games are an example of \textit{asynchronous multiplayer}, sometimes referred to as \textit{mingleplayer}, in which players feel the impact and presence of other players without direct live interaction between them. Since Demon's Souls, players have been able to leave messages on the ground using fixed templates and a limited vocabulary, which are then sent to a game server. When a player enters an area, messages from other players are randomly retrieved from the server and placed in the world exactly where they were written. This messaging system has been used by players to warn about dangers, guide to secrets, make jokes, trick players into death or danger, and more.

Soulsborne games have two other related features: bloodstains, and player ghosts. Whenever a player dies, the last few seconds of their player's actions are recorded and sent to a game server. When a player enters an area, there is a chance that they will see bloodstains on the floor which mark the beginning of a player death recording. If they interact with the bloodstain a red ghost of the dead player will appear and play out the last moments of their life. Only the dead player's actions are shown - any effects, interactions, items, spells or enemies are not recorded.  Ghosts are non-interactive and do not affect the current player's world. Similar to in-game messages, ghosts provide information to the player. They may show other players being ambushed, being caught by hidden traps, opening secret doors, or using particular strategies.

The messages in Elden Ring have garnered a large amount of interest from games journalists due to their propagation of memes and their often deliberately misleading nature. While some journalists believe that the messages make the game feel like a conversation and enjoy the in-jokes \cite{marshall2022} \cite{thorn2022}, they have also been criticised as a distraction to gameplay, enabling toxic players \cite{park2022}. An article by Cian Maher \cite{maher2022} elaborates on how jokes and memes are often lost in translation, such as ‘fort, knight’ (a reference to the Battle Royale \textit{Fortnite}) being misinterpreted by Japanese players assuming they referred to a special event happening at night.  Bloodstains have not received the same critical attention, though their potential humorous nature has been highlighted by Kapron \cite{kapron2022}. They discuss how a video was shared on Reddit by the user Shinokijorainokage, touching four bloodstains to reveal different players jumping off the same balcony for no apparent reason. This example demonstrates how bloodstains add to the gameplay experience of Elden Ring and create an anonymous camaraderie among players.

\subsection{Archaeological Surveys in Games}


Video games have been considered by archaeologists as artefacts of study for several decades \cite{mol2017}, but the term ‘archaeogaming’ was not coined until 2013 \cite{reinhard2013}. Archaeogaming research broadly covers the ethical representation of heritage in the medium \cite{dennis2019}, creating games from an archaeological perspective\cite{copplestone2017} and treating video games as archaeological sites \cite{reinhard2018}. More recently, there has been collaboration between computer science and archaeological practitioners to contextualise early game development tools \cite{aycock2020}.

To date there have been a very limited number of archaeological surveys in digital games. The most significant was Reinhard’s \textit{No Man’s Sky} Archaeological Survey conducted in 2018 \cite{reinhard2019data}. The player community known as the Galactic Hub was displaced by the Atlus Rises update in 2017 which changed the nature and topography of planets they inhabited \cite{reinhard2021}. In anticipation of the NEXT software update in 2018, Reinhard used various methods to record Galactic Hub player settlements, including maps, screenshots, photography, videography, excavation and field reports. 

Smith Nicholls \cite{smithnicholls2018} conducted an archaeological survey of player deaths in \textit{NieR:Automatata}. The game has a similar asynchronous multiplayer system as the Dark Souls games called the reliquary system, which displays other player’s corpses on the map. The survey was experimental in nature, mapping corpse locations on a map as part of a desk-based assessment of the site  \cite{cifa2014}, and also formed part of a study of how archaeogamers have used mapping techniques to record games \cite{smithnicholls2021}.

Although not explicitly an archaeological survey, Graham \cite{graham2020} recorded a specific playthrough of \textit{Minecraft} from an archaeological perspective through text vignettes and screenshots, emphasising the ethical implications. 

\subsection{Elden Ring as an Archaeological Site}

There is a considerable precedent for understanding the Soulsborne games from an archaeological perspective. In 2012, Forbes published an article titled `The Wonderful Archaeology Of `Dark Souls' Lore’ \cite{kain2012} which delves into how the game excels at environmental storytelling, requiring the player to interpret the narrative through its ruined world. This is a point also emphasised in a 2017 \textit{Eurogamer} article by archaeologist Philip Boyes \cite{boyes2017} who comments that ``it invites the player to share in the narrative process and become a researcher-cum-author themselves.'' Focusing on Bloodbourne specifically, Kerry Todd has also recently published an article on the narrative archaeology of the game \cite{todd2021}.

More recently, archaeologist Bill Farley has covered the Soulsborne games in his YouTube series \textit{Video Game Archaeology}. In one video \cite{farley2021} he uses the Undead Asylum in Dark Souls to explain archaeological theory, and he has also produced commentary videos about Elden Ring from an archaeologist’s perspective \cite{farley2022}.
 
As detailed above, Elden Ring is a hugely popular game with mingleplayer features that allow the player to interact with both messages made deliberately, and death records made incidentally, by other players. If we understand that video games can be studied as archaeological sites, then Elden Ring presents an exciting case study given the wealth of player data available to us at the time of writing through these features. As a team, we represent expertise in both archaeogaming and computer science. This collaboration is important, as Aycock puts it: “archaeology and computer science are full-time pursuits, and it is only by combining forces in an interdisciplinary effort that these digital artefacts can be thoroughly examined” \cite{aycock2021}.


\section{Methodology} \label{sec:methodology}
Building on previous work archaeologically recording player-created content in video games, this survey aimed to record a sample of player message and bloodstains in Elden Ring. As stated in the background section, we wanted to record this sample in March 2022 as it represented a period of high play activity in the game following its initial release in February 2022. As a two person team consisting of an archaeologist and computer scientist, we intended to combine our expertise to undertake this project. As such, we devised the following research questions:

\begin{itemize}
  \item Is it possible to archaeologically record player messages and bloodstains in Elden Ring?
  \item What do the messages and bloodstains indicate about player experience of Elden Ring?
  \item How do these messages or bloodstains inform us about the metagame of Elden Ring?
  \item How are the results of this study applicable to game design and archaeogaming research more broadly?
\end{itemize}

\subsection{Code of Ethics}
As Shawn Graham states “Video games are built environments and thereby invite archaeological study, in which case professional archaeological ethics should apply” \cite{graham2020}. Meghan Dennis has done crucial work on the ethics of archaeogaming, which has been instructive for this investigation. We draw attention to her doctoral thesis \cite{dennis2019} and her 2016 article ‘Archaeogaming, Ethics, and Participatory Standards’ \cite{dennis2016}. The section on researching in a multiplayer environment is especially relevant to this study, as we had to consider our roles as surveyors in the game world as well as our duty of care towards other players. Unlike other online multiplayer games, Elden Ring does not provide any extraneous information regarding the identities of players whose messages and bloodstains you encounter, such as aliases. Regardless, it is imperative that we treat all data we collect in Elden Ring with respect and in full knowledge that it was created by other participants in the player community.

Another key source for our study is the No Man’s Sky Archaeological Survey Code of Ethics\cite{flick2017}, as this presents an example of how an archaeological code of ethics was co-created for a similar project. We highlight Principle 5 of the Code in particular:
 “Ensure the integrity of archaeological sites, humans and non-human people and animals, and archaeological artefacts where possible; work to ensure
good stewardship of sites, peoples, and artefacts; and avoid and discourage
activities that enhance the commercial value of archaeological artefacts” \cite{flick2017}. 
As Dennis has demonstrated in her aforementioned thesis \cite{dennis2019}, archaeological artefacts, if they are present in video games, are often commodified through looting mechanics. This reflects real-world unethical looting practises that damage archaeological sites for commercial gain. A key guideline for our work is to consider the nature of our survey and to ensure our methodology does not contribute to the commodification of the archaeological record, whether digital or analogue.

Another aspect to consider in our Code of Ethics was Elden Ring’s own 'Manners During Online Mulitplayer.' In order to start the game, the player must agree to this code of conduct. As can be seen in Figure 3, one of the requirements of this code is “Do not play games by using the in-game functions in a manner other than what they were originally intended for” \cite{fromsoftware2022}. This general statement could impinge upon our survey -- after all, in-game messages and bloodstains were certainly not intended to be studied archaeologically. However, Graham’s point that “If we play games not as ethically informed archaeologists, if we do not write about games or critique games from an archaeological perspective, we are submitting to the power of the game publisher and the game maker to set the terms of reference about the past” \cite{graham2020}. Even if archaeologically recording Elden Ring does not technically fall within the intended player experience of the game, it is a valuable undertaking in providing that record.

\begin{figure}[t] 

\includegraphics[width=\columnwidth]{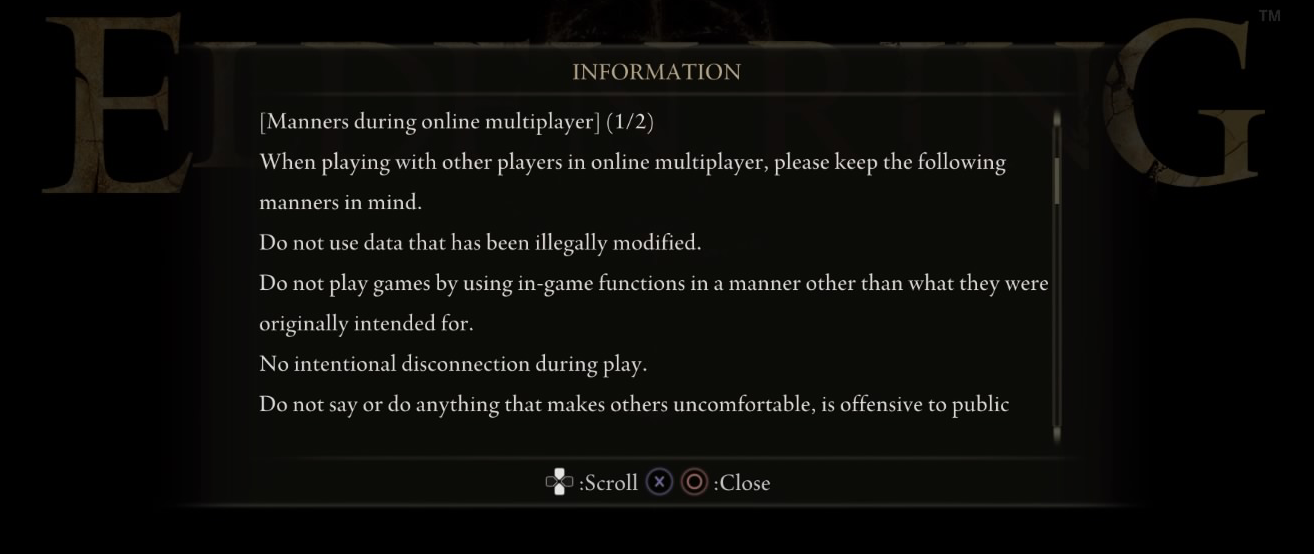}
\caption{Elden's Ring's 'Manners During Online Multiplayer.'}

\label{fig:screenshot}

\end{figure}

We follow all appropriate guidelines as have been set out in the No Man’s Sky Archaeological Survey Code of Ethics\cite{flick2017} and the Chartered Institute for Archaeology’s Code of Conduct \cite{cifa2014} and the Computer Applications \& Quantitative Methods in Archaeology Ethics Policy \cite{CAA2021}. We have also drafted our own Elden Ring Survey Code of Conduct Guidelines to accommodate the specific requirements of our study, as follows:
\begin{enumerate}
  \item Survey members will not rate any messages they encounter. Survey members are not to impose or express any value judgements on any messages, bloodstains or any other aspect of player data they encounter.
  \item Survey members will not leave any of their own messages in the survey area. 
  \item Survey members will not transcribe any messages which they deem to contain derogatory content or have derogatory connotations.
  \item Survey members are to take their role as stewards of player data seriously. In particular, blood stains show player deaths, which are understood to be potentially vulnerable moments for players.
  \item Survey members will uphold professional archaeological standards and maintain accuracy in their recording as far as reasonably possible. It is understood that network connectivity or other technical issues could interfere with surveying. If this is the case, survey members will make a note of any such interruptions. 
  \item Survey members are to put their own health, safety and wellbeing first. The archaeological surveying of video games can be intense work. Survey members are to take frequent breaks during the recording process.
  \item The results of this survey will be publicly accessible, published on the authors' Research Collective website.
\end{enumerate}

\subsection{Site Selection}

We initially considered a large open area in Elden Ring and modelled our investigation as a kind of fieldwalking survey \cite{fieldwalking}. We chose an area in Limgrave (see Figure 4) with varied topography and enemies. Limgrave is the first area that the player encounters after the initial prologue of the game, and is also an area that both surveyors were familiar with, so it presented an appropriate case study. As part of a trial study, we mapped the location of enemies and other points of interest on the map using in-game markers. Once we tried to record the location of player bloodstains, however, it became clear that using the in-game map was too schematic for this purpose. Moreover, the density of bloodstains that would appear over time meant even within one small building (the Church of Elleh) there was a large amount of data to record. For this reason, we decided instead to focus our attention on two small self-contained areas in Limgrave.

\subsubsection{Church of Elleh} \label{sec:ellehsite}
\begin{figure}[t] 

\includegraphics[width=0.9\columnwidth]{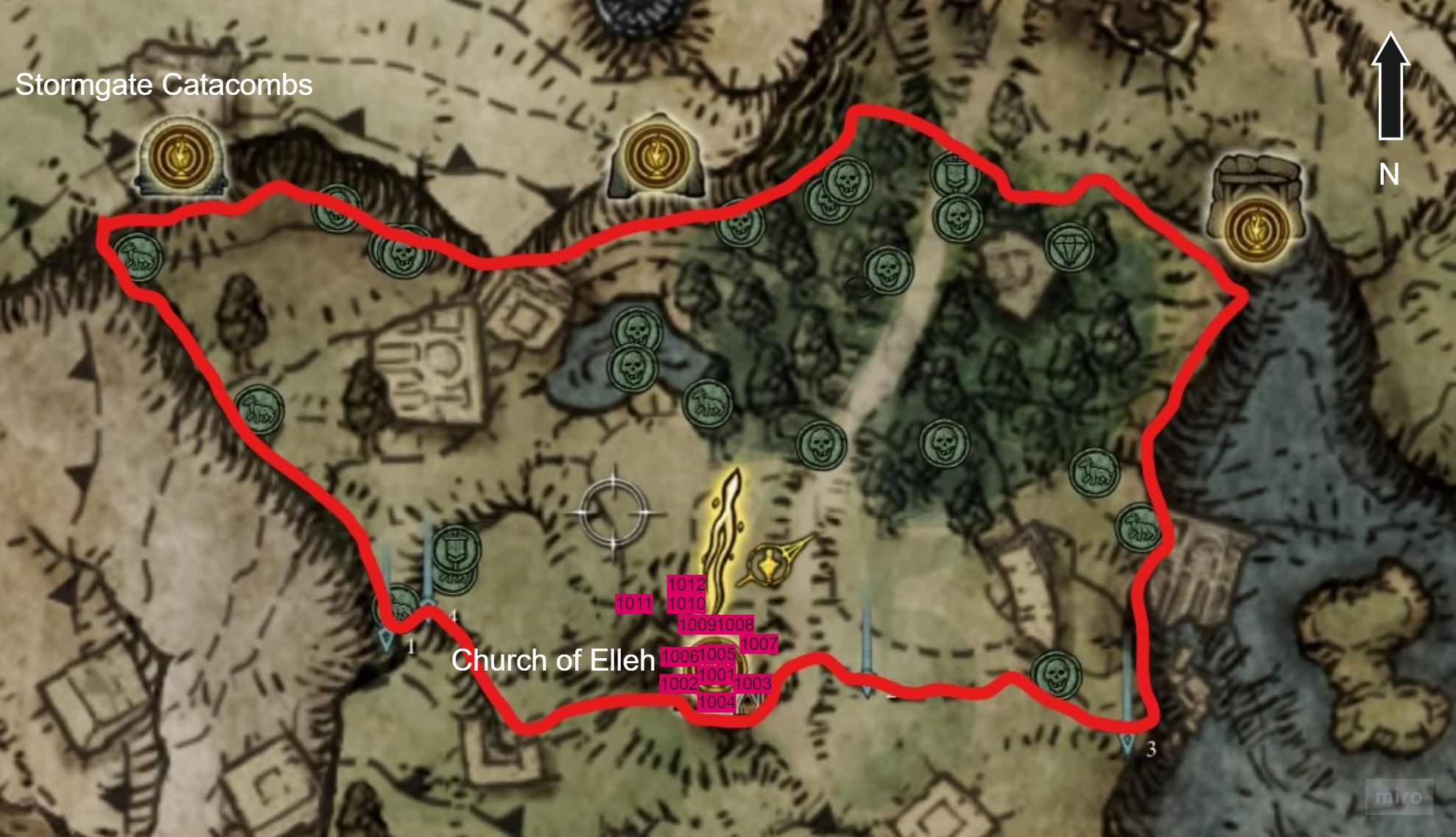}

\caption{Map of trial study area outlined in red. Bloodstain locations are denoted by numbers highlighted in red.}

\label{fig:map}

\end{figure}

The Church of Elleh is a ruined church located due north of the location the player begins the game in. The player is explicitly led to the Church as part of the game's critical path by following golden trails of light, and corresponding golden map markers (one of which can be seen in Figure \ref{fig:map}, guiding the player north from the church). The Church contains two important features: an anvil, which allows players to upgrade their weapons (the only way to do this during the first few hours of the game); and a merchant, which allows players to purchase items (likely the first merchant the player encounters, although it is possible to meet others by exploring off the main path).

New players are also likely to encounter two key narrative beats in the church, too. Melina, an NPC who is central to the plot and also provides players the ability to level up and increase their strength, can appear for the first time at the Church of Elleh\footnote{This is slightly variable as it triggers based on player exploration, but Elleh is a common site to encounter Melina in.}; and Ranni, an NPC who has a large secondary storyline and quest, also appears to the player here and provides access to a rideable animal\footnote{Ranni only appears at night, thus it is possible but unlikely to miss this.}. 

In addition to this, the Church is close to optional bosses, dungeons and other special game content, many of which the player will be encountering for the first time. Therefore, the Church represents an excellent location to study, as it has narrative, mechanical and structural significance for the player, and thus should allow us to observe a rich collection of player experiences and recorded reactions. We chose to survey all of the ground within the boundaries of the church walls.

\subsubsection{Stormgate Catacombs}
Elden Ring's world contains many caves, catacombs and dungeons which act as optional side content for players. Most of them follow a similar pattern: a linear sequence of rooms, often with a unifying mechanical theme, culminating in a boss fight with a significant reward. Stormgate Catacombs is geographically one of the closest dungeons to the start of the game, however it is not necessarily the most commonly encountered dungeon as it is off the main path and requires exploration to find.

We selected Stormgate specifically as it is one of the earliest examples of a side area with both novel mechanics and ambush traps. Stormgate contains several rooms in which enemies are hiding behind entrances waiting to flank the player as they enter, and it is also the first appearance of `fire pillars', stone statues that periodically shoot fire down corridors. Attacking the pillars in any way causes them to recess into the floor and stop shooting fire.

A key role for both bloodstains and messages in areas such as this is to act as a warning for players. Bloodstains indicate danger and can give hints as to how someone died, and messages are often used explicitly to warn others. We theorised that Stormgate Catacombs would have good examples of both of these, given the nature of the traps and ambushes present. We chose to survey a corridor midway through the catacombs that immediately precedes the room containing the first fire pillar, and an ambushing enemy.

\begin{figure}[t] 

\includegraphics[width=\columnwidth]{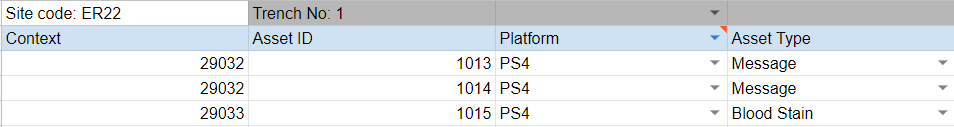}
\caption{Extract from survey spreadsheet}

\label{fig:spreadsheet}

\end{figure}

\subsection{Data Collection}
When considering methodologies for data collection, it was important to prioritise what was most relevant to our research questions. Being able to identify and record individual player messages and bloodstains was our focus, so we considered both as artefacts and created a spreadsheet modelled on an archaeological small finds register \cite{archaeologicalresearchservices2018}. Unlike artefacts in the analogue world, messages and bloodstains are all of identical dimensions and usually at the same depth on the ground (edge cases are messages which are placed on in-game surfaces). Each artefact was given a unique ID corresponding with the ‘trench’ it was found in. Each surveyor played the game on a different platform (PC and PlayStation4) which was also recorded in the spreadsheet for context. In addition, there were fields for ‘Associated Game Assets’ and ‘Associated Player Created Assets’ to capture if messages or bloodstains were in close proximity to game assets such as NPCs, or if they were associated with other player created messages and bloodstains.

The ephemeral, fleeting nature of both bloodstains and messages in the game presented a challenge, as they could appear and disappear within a few minutes depending on when the game server updated. We approached this challenge through the lens of the archaeological context. An archaeological context is the “position and associations of an artifact, feature, or archaeological find in space and time. Noting where the artifact was found and what was around it assists archaeologists in determining chronology and interpreting function and significance” \cite{archaeologicalinstituteofamerica2022}. While we recorded the spatial context of each message and bloodstain on a plan (see the section below) we recorded the temporal position of each asset relative to when it was observed on a particular day and at what observable number of server refreshes. Figure 5 provides an example of this from our spreadsheet. In this example all three assets (1013, 1014 and 1015) were recorded on the same day (29th of March), but 1015 appeared after the other two and so is given the context number 29033.

\subsection{Archaeological Planning}

Recording the location of messages and bloodstains relative to bounds of the surveyed areas also presented a challenge for our study. In the case of Reinhard’s No Man’s Sky Archaeological Survey, he used time maps (a record of how long it took to walk to different assets from a set point) and aerial photography to make plans. Aerial photography is not an option in Elden Ring, and our survey ‘trenches’ were small enough that scaled plans were deemed more appropriate. Archaeological plans are a standard method for recording the extent and surface of a context \cite{MOLAS1994}. Using an online graph paper tool, we were able to create base plans of the Church of Elleh (Figure 6) and the Stormgate Catacombs (Figure 7) by using a player avatar 'tarnished foot’ as a measurement.

This method was time-consuming and required some approximations, but it did produce base plans that we could use for the purpose of our survey. Where possible, in-game assets were recorded on the base plans. In the case of the Stormgate Catacombs plan, for example, only skulls and not individual bones were recorded on the plan as there was some concern this might make the recorded messages and bloodstains less legible. To our knowledge, this is the first time that hand-created plans to scale have been used as part of an archaeological survey of a video game, rather than in-game maps, photographs or more abstract methods. 

\subsection{Photography}
As in analogue archaeology, photography was a crucial recording method in our archaeological survey. Initially we also considered recording video footage, but decided that for the purpose of this study screenshots would be sufficient to record the location and contents of messages and bloodstains. On further reflection, video footage would have been particularly useful as an aid in analysing bloodstain ghosts that often moved quickly and were difficult to capture with screenshots.
Elden Ring does not have a dedicated photo mode. There is a mod available which allows players to move the camera around freely and pause the game \cite{colp2022}, however it has to be used offline, rendering it useless for our survey. 
Archaeological photos usually contain some form of scale for later reference. In our case, our own player avatars functioned as a scale in the screenshots. Though this was effective for the purposes of our survey, this practice of using a human body as a scale in archaeological photography does have colonial antecedents \cite{chadha2002} and we would aim to explore alternate methods in future, especially in other games.

\begin{figure}[t] 

\includegraphics[width=0.9\columnwidth]{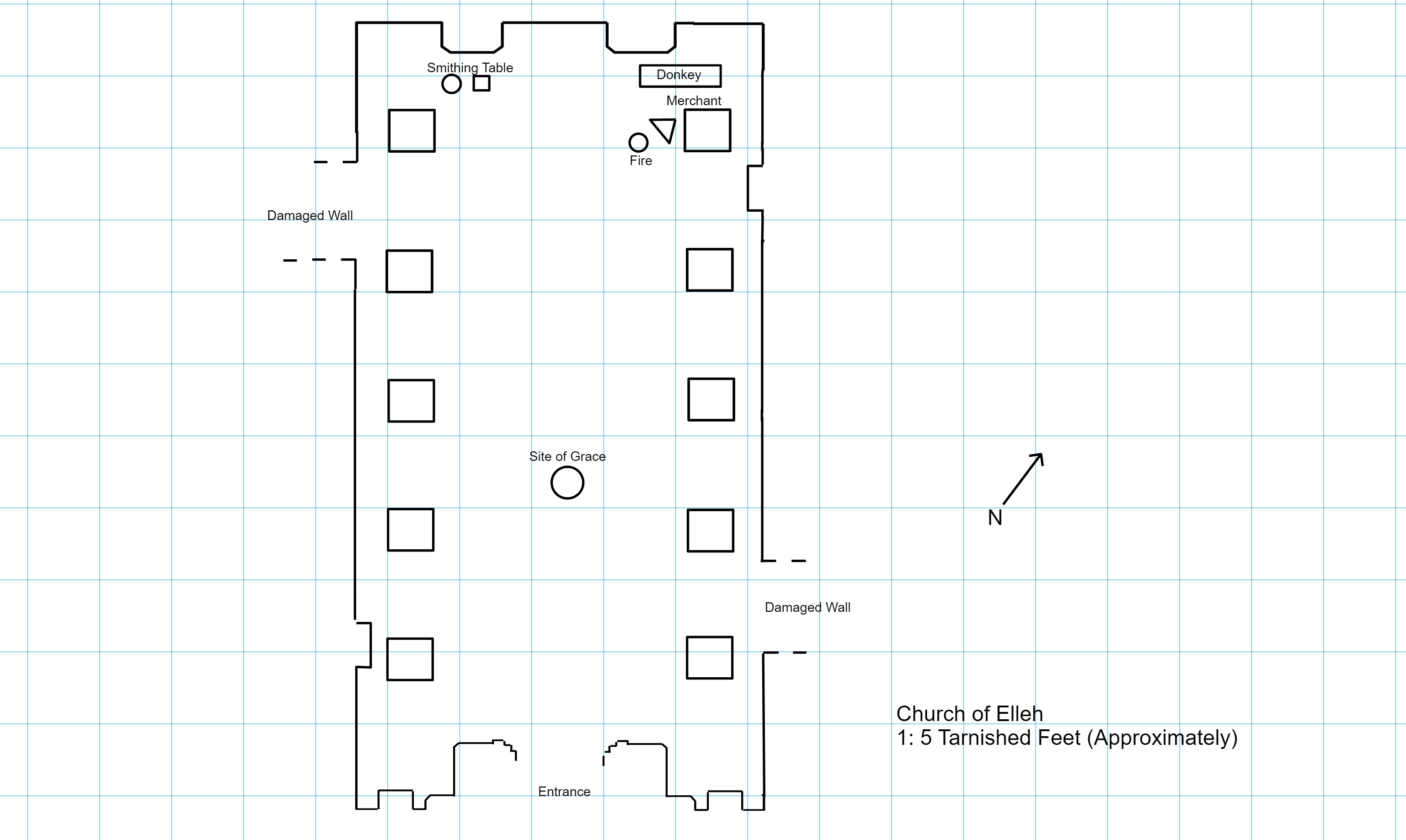}

\caption{Church of Elleh Base Plan}

\label{fig:church}

\end{figure}

\begin{figure}[t] 

\includegraphics[width=0.9\columnwidth]{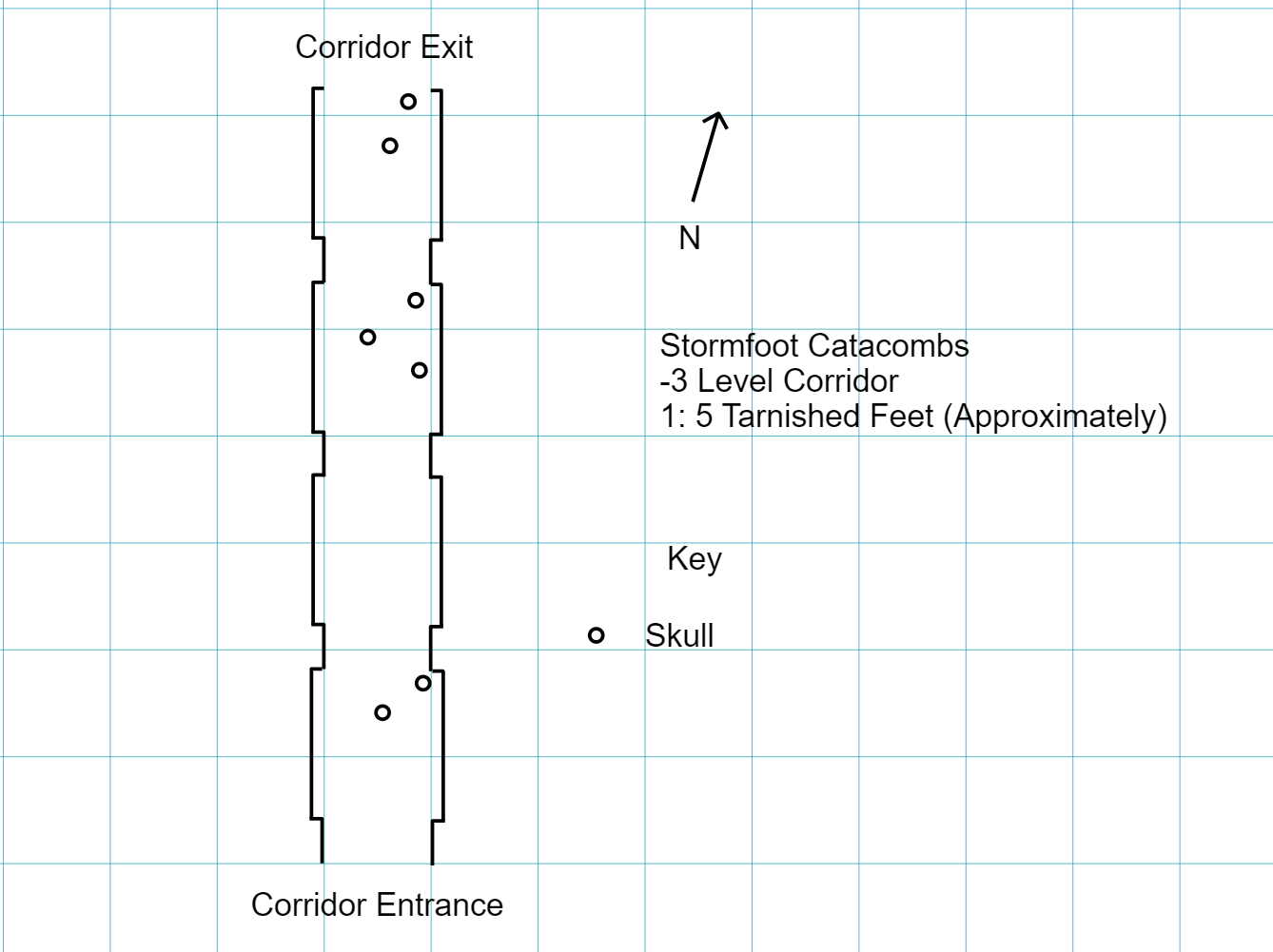}

\caption{Stormgate Catacombs Base Plan}

\label{fig:catacombs}

\end{figure}

\section{Results} \label{sec:results}
Not including the pilot study, the survey was undertaken between the 29th and the 31st of March 2022. In total, 14 bloodstains and 14 messages were recorded in the Church of Elleh, and 8 bloodstains and 4 messages were recorded in the Stormgate Catacombs. Each location was surveyed on three discrete occasions (once on the 29th by each surveyor and once on the 31st by one surveyor). The greater number of records produced in the Church of Elleh likely reflects that location serving as a player hub and it representing a larger surveyed area. The results of the survey will be discussed by location and context.

\subsection{Church of Elleh}
\subsubsection{Contexts 29032 and 29033}

Both contexts were recorded on the PS4. Context 29032 contained two messages (1013 and 1014). Message 1013 read ‘Praise the fingers!” which is assumed to be a reference to the Two Fingers NPC. Message 1014 reads ‘Behold, something incredible!’ and was placed in front of the smithing table, perhaps as a joke. Context 29033 represented a discernable change with the appearance of four bloodstains which were rapidly batch recorded under the asset ID 1015 in the northern half of the church. Unfortunately, this meant that no further data pertaining to the bloodstain ghosts were recorded in this case.
\subsubsection{Context 29036}
Three messages were recorded within this context, all located near the Point of Grace within the church. Message 1016 read ``Seek lover'', and was placed directly in front of where the NPC Ranni appears, as described in Section \ref{sec:ellehsite}. Such messages serve a dual purpose: to players who are yet to encounter Ranni they act as player-authored foreshadowing, whereas experienced players may experience them as the textual equivalent of a knowing nod, acknowledging the shared experience. Message 1017 and 1018 were closer to the Point of Grace itself. The first reads ``stay calm/skeleton''. We interpret this as addressing the player as skeleton. In previous Dark Souls games the player is explicitly `undead' in the narrative - in Elden Ring they are referred to as `Tarnished' but similarly unable to die permanently. Addressing the player as skeleton would make sense. Message 1018 reads ``First off, safety/And then time for lover''. This is an instructive message to new players, informing them this is a safe place, and following up with foreshadowing a meeting with either Melina or Ranni. Female-coded NPCs in From Software games are often referred to romantically in messages, and the NPC calling herself Ranni in particular became a fan favourite shortly after release.

\subsubsection{Context 29037}
Shortly after recording Context 29036, walking across the site to the merchant, two messages were spotted. Message 1019 read ``Be wary of dog'', and is located just in front of the donkey belonging to the merchant. A consistent joke throughout Elden Ring's community is mislabelling any non-canine animal as a dog, which this is a prime example of\footnote{Later in the game, many canine animals are found next to messages that label them as anything \textit{but} dogs.}. Message 1020 expired while recording Message 1019, so its content is not known to us, but its location is significant: it is written directly behind the donkey's hind end. Given our experience of community messages throughout Elden Ring, we suspect this message was probably a joke at the donkey's expense.

\subsubsection{Context 29038}
Turning around from Context 29037, three bloodstains were spotted -- two close to the merchant, with a third further away near the point of grace. Bloodstain 1021 shows a character fighting near the merchant, wearing armour that is found in mid-game areas after advancing certain questlines. This is notable as the Church is very near the start of the game. Bloodstain 1022 is located almost on top of the merchant, and may depict a player fighting the merchant and dying. They also appear to be wielding a lategame weapon, in this case one obtained by fighting a secret boss at the very end of the game. Bloodstain 1023 is located nearer to the Point of Grace, close to Message 1018, and thus is further away from the merchant so the cause of death is harder to identify. The player wields another late game weapon found in a rare area. 

\subsubsection{Context 31031}

This context was recorded on the PS4 and contained three messages, two of which (1025 and 1026) appear to be identical to 1013 and 1014. This indicates that it is possible for messages not to be refreshed in an area, even after several days have elapsed. Message 1024 was located close to the Site of Grace and read “Time for night.” This is assumed to refer to the ability to allow time to pass at a Site of Grace. It could potentially be referencing the presence of the NPC Ranni the Witch, who appears at the Church of Elleh on one occasion at nightfall. 

\subsubsection{Context 31032}
Context 31032 was also recorded on the PS4 and contained four bloodstains (1027-30). All of these bloodstains showed players dying in combat.  Bloodstain 1029 was particularly interesting as it displayed a player wearing the Iron Kasa helm. This helm can only be acquired be following a specific NPC questline \cite{eldenringwiki2022}, indicating this player had completed this and returned to the Church afterwards. 

\subsubsection{Context 31033}
Context 31033 was also recorded on the PS4 and contained eight bloodstains and two messages. This context rapidly appeared after 31031 and 31032, indicating the speed with which the player created content can change the experience of a space in the game. Out of these bloodstains, 1034 and 1035 are of particular note. Bloodstain 1034 showed a player holding the Fingerprint Stone Shield. This shield can only be obtained after defeating a lategame boss and dispelling an illusory wall \cite{webb2022}. It is a desirable item as it is one of the best defensive armaments in the game. Bloodstain 1035 also showed a particularly desirable item: the Sacred Relic Sword. This sword can only be obtained the game by trading an item which can only be obtained by defeating the final boss of the game, indicating that this player may be on their second playthrough of the game having carried over items obtained in their first playthrough.
Messages 31033 and 31044 were both located near the entrance to the Church of Elleh and appear to be earnest guidance to other players. Message 31033 read “merchant ahead and then sorcerer ahead” and message 31044 read “friend ahead.” 

\subsubsection{Context 31034}
This context was recorded on PS4 and occurred as a result of the surveyor accidentally dying at the hands of the Tree Sentinel by the entrance to the Church of Elleh. Upon death, the surveyor observed the Tree Sentinel damaging the church itself. This observation was confirmed to be true by a Reddit forum post \cite{darkcrowi2022}. Rather fittingly, message 31034 was the only asset recorded in this context upon the surveyor respawning and returning to the entrance, and it read “First off, keep moving” which was likely a reference to the danger of the Tree Sentinel.

\subsubsection{Context 31035}
This context was the last recorded on PS4 in the Church of Elleh and contained only one message, 1040, which read “Could this be a love?” Like message 1016, this likely refers to the NPC Ranni the Witch.

\subsection{Stormgate Catacombs}

\subsubsection{Context 29031}
This context was recorded on the PS4 and contained only one bloodstain (2001), which reflects the paucity of player created assets that could on occasion spawn in this area. The message was located at the end of the corridor and likely was a result of the player dying to the Imp or Fire Pillar in the next chamber.

\subsubsection{Context 29034}
In this context two messages were discovered at the end of the corridor, immediately preceding the opening into the next room. Message 2002 reads ``Be wary of up'', a reference to the Imp waiting to pounce on the player, clinging to the wall above and to the right of the doorway. Message 2003 reads ``fire ahead'', a reference to the fire pillar trap in the next room. Both messages act as warnings for the player about obstacles ahead.

\subsubsection{Context 29035}
This context was encountered outside of the official site area, in the room directly following the corridor. Two bloodstains and one message were observed and recorded. Message 2005 reads ``Be wary of fire/Try attacking''. This is a cryptic reference to the fact that fire pillars will deactivate if hit by an attack, therefore a ranged attack fired down the corridor will render it completely safe. Bloodstain 2006 shows a player in mage's robes fighting something (most likely the imp) and dying, while bloodstain 2007 shows a player running quickly down the corridor and then dying, almost certainly to the fire. The fire pillar periodically stops to allow players time to dash, this player mistimed their run and died just inches short of the next safe zone.

\subsubsection{Context 29036}
This context was recorded on the PS4 and contained one message (2008) and one bloodstain (2009). Message 2008 read “be wary of fire” which is assumed to be a reference to the fire pillar in the next chamber. The bloodstain showed a player of likely Samurai class running from the south-east, potentially attempting to run away from one of the Imps in the preceding chamber.







\section{Discussion} \label{sec:discussion}

\subsection{Surveying Elden Ring}
This study represents one of only a handful of videogame archaeological surveys that have ever been undertaken, and the only known archaeological survey of any Soulsborne game. Furthermore, this is the only known study which used hand-drawn plans to scale as part of the recording process. The survey was small with only 40 records made in total (not counting the pilot survey) and would certainly not be applicable for quantitative study. However, we believe it does represent a successful qualitative sample of the asynchronous mingleplayer experience in the game, proving that it is possible to archaeologically record player messages and bloodstains in Elden Ring. Moreover, and more importantly, we were able to respectfully and accurately record player messages and bloodstains in line with our stated Code of Ethics.

The experience of creating the base maps was intense screen-based work, and we would certainly flag this as an important wellbeing consideration for future surveys. We would hope to see more engagement with creative plan-making in archaeological surveys of videogames. A recent study \cite{morgan2021} highlighted the importance of drawing for archaeologists in terms of forming mental maps, and our experience creating plans of the Church of Elleh and Stormgate Catacombs confirmed this. For example, the in-game experience of carefully delineating each survey area aided us in understanding how the spaces were constructed from repeated assets.

\subsection{Excavating the Meta}
One of our proposed research questions related to the possibility of recording aspects of the Elden Ring metagame. A metagame can be defined as a game within a game, in which players create their own rules beyond the ‘canon’ constraints of the original. The player messages in Elden Ring arguably constitute a metagame in themselves as they represent a broader tradition of players using the feature in Soulsborne games to make jokes and reference memes propagated through social media, as mentioned in our background section.
In the Church of Elleh, we identified three potential references to the NPC Ranni the Witch which referred to her as ‘love’ or ‘lover’  (1016, 1018, and 1040). We were able to theorise this based on knowledge of the Elden Ring fandom, in which Ranni is a fan favourite character. This is reflected in reporting on fanart of the character \cite{mcnulty2022}\cite{warren2022}. These messages represent a kind of metagame in encouraging or guiding other players to the character in-game. In addition, message 1019 is a classic example of the Elden Ring meme of referring to any in-game animal as a dog. 

In considering our pursuit of the metagame, we found this point from Patrick LeMieux and Stephanie Boluk’s Metagaming \cite{boluk2017} to be particularly instructive:
``The concept has taken on renewed importance and political urgency in a media landscape in which videogames not only colonize and enclose the very concept of games, play and leisure but ideologically conflate the creativity, criticality, and craft of play with the act of consumption.''
In the Code of Ethics section we discussed how this survey could be interpreted as falling outside of the Elden Ring ‘Manners During Online Multiplayer.’ In a sense, this survey was a metagame, requiring the surveyors to engage with the gamespace creatively in order to create an archaeological record. That being said, by highlighting aspects of online fan culture in our analysis of the game, perhaps we are falling into the trap of conflating creativity with consumption in exactly the way that LeMieux and Boluk warn against.

\subsection{`Fashion Souls' as Archaeology}

‘Fashion Souls’ is a specific subset of the Elden Ring player community who share outfits they put together in the game, mainly on a subreddit boasting over 80,000 members \cite{reddit2022}. So influential is this digital sartorial community that they were even featured in a mainstream fashion magazine  \cite{gordan2022}. This aspect of player culture and experience became relevant to our survey when we realised that we could identify specific armour and weapons from the bloodstain ghosts. For example, when recording bloodstains in the Church of Elleh we found some player ghosts wearing starter clothing for certain character classes. This makes sense, as the Church is one of the first locations a player will encounter in the overworld. Other bloodstains, however, wore more unusual armour sets and weapons. We were able to identify some of these through our own experience of playing the game, while others required looking through the player-authored wikis and comparing our screenshots with online images. One weapon in particular we sought out identification by consulting an Elden Ring expert player. 
This aspect of our analysis was particularly interesting because it mirrors the practice of consulting with finds specialists in analogue archaeology, such as experts in pottery or metal artefacts. Our observations on players with items or armour from the late game of Elden Ring also has implications for studying player experience in demonstrating that the Church of Elleh is a hub that players return to, despite being in the starting area of Limgrave.

\subsection {Elden Ring as Palimpsest}
The metaphor of the ‘palimpsest’ is perhaps overused in archaeology, but in this case we believe it is highly appropriate. A palimpsest is a manuscript which has been scraped or washed so that it can later be re-used, often with fragments of the original writing remaining. With the continual overlay and removal of player messages, Elden Ring can be considered as a digital palimpsest. Bailey \cite{Bailey2007} has discussed the application of the palimpsest metaphor in archaeology. The Elden Ring messages are potentially an example of what he defines as true palimpsests in the sense that all remains of previous messages will eventually be removed completely when the server updates for a specific player. However, as messages often reference each other, there is a sense in which partial impressions of other players’ experiences are also preserved.
Bailey argues that palimpsests, in potentially totally erasing or obscuring earlier contexts, present a challenge for archaeologists. In particular, he argues that:
``we cannot work out what tools we need until we know what sort of phenomena are there in the longer-term record to investigate, and we cannot investigate those different phenomena until we have some tools to do it with.''
This paradox is at the heart of archaeogaming research -- we need to develop new tools and methods for the archaeological recording of immaterial space, but we cannot know what methods will be most effective until we apply them. This survey constitutes one potential approach for recording complex digital palimpsests, and we hope to build on this in future.

\section{Future Work} \label{sec:futurework}
\subsection{Mechanised Surveying}
Different types of archaeological survey and site present their own sets of opportunities and affordances, and digital worlds are no different. Games provide an opportunity to use other software to enhance or automate some parts of the archaeological survey process. For example, both messages and bloodstains must be retrieved from the server before they can appear in-game. In theory, we could watch network traffic and (assuming the information was not encrypted) automatically collect this information from the server, instead of finding it in-game. However, there would be ethical concerns about harvesting player data using this approach, and it would definitely contravene Elden Ring's Online Code of Conduct.

This approach would also have many implications for the nature of the survey. We gained many important insights by experiencing the data \textit{in situ}, as well as a better understanding of the environment itself. It would also prevent us from experiencing the content in the way another player would. Despite these shortcomings, we are interested in employing computer science techniques to future survey work. For example, automation could be used to provide a preliminary assessment of a larger area to help select a smaller site to survey manually, perhaps by measuring the density of artifacts in a particular location.

\subsection{Longitudinal Surveying}
It is unclear to us, at the time of writing, how the distribution of messages in a From Software game changes over the lifespan of the game. Dark Souls was originally released in 2011, and few people play the game regularly today compared to its peak. Yet playing it today still shows player ghosts, bloodstains and messages, as the servers retrieve older messages and display them. Playing Elden Ring in 2027 or 2032 will likely yield a different mix of messages and player ghosts to today -- although some of the messages we have recorded may still be seen by some players.

Performing several studies of this kind, interspersed over a long period of time, could help yield new insights into how the composition of messages and ghosts changes over time. This could show us evidence of Elden Ring's social meta developing, the composition of the playerbase changing (since more hardcore players tend to make up a larger proportion of the community as the game ages), and the nature of player communication maturing.

\section{Conclusions} \label{sec:conclusions}
In this paper we present a proof of concept for archaeologically surveying ephemeral player-generated content in games such as Elden Ring. To date, there have been an extremely limited number of archaeological surveys of videogames. This work helps fill the gap in this fledgling research area with new recording techniques and methodological approaches. Through a survey of just two sites, we surveyed 40 messages and bloodstains, which represented a rich qualitative sample of two areas in the game. Furthermore, this work included the first known use of scale plans to archaeologically record a videogame. 

Our findings show that archaeological surveys are an effective method not just for gathering user-generated content directly, but also for observing the culture and essence of a player community. Player messages referring to the NPC Ranni, for example, are indicative of a wider metagame of using messages to signpost and comment on popular characters, events and mechanics. It was also possible to make specific inferences about player behaviour through observing player avatars -- we were able to interpret the armour worn by bloodstain ghosts to deduce that even after accessing late game content players return to the Church of Elleh. This shows that this research has applications in the study of player behaviour and experience, as well as archaeogaming and game design. 

\section{Acknowledgements}
The authors wish to thank Peter White and Mark Quane for providing expert advice on Elden Ring, and to Dr. Brianna McHorse for her advice on equine terminology.



\bibliographystyle{ACM-Reference-Format}
\bibliography{bibliography}

\end{document}